\documentclass[twocolumn]{webofc}

\usepackage[varg]{txfonts}   % Web of Conferences font
\usepackage{hyperref}
\usepackage{url}
\hypersetup{colorlinks=true,citecolor=blue,urlcolor=blue,linkcolor=blue}
\begin{document}
\title{Muon Nuclear Data Development Project}

\author{\firstname{Yukinobu} \lastname{Watanabe}\inst{1,6}\fnsep\thanks{\email{yukinobu.watanabe@kyudai.jp}}
        \and
        \firstname{Megumi} \lastname{Niikura}\inst{2}
        \and
        \firstname{Shinichiro} \lastname{Abe}\inst{3}
        \and
        \firstname{Sayani} \lastname{Biswas}\inst{4}
        \and
        \firstname{Hiroki} \lastname{Iwamoto}\inst{3}
        \and
        \firstname{Adrian} \lastname{Hillier}\inst{4}
        \and
        \firstname{Naritoshi} \lastname{Kawamura}\inst{5}
        \and
        \firstname{Shoichiro} \lastname{Kawase}\inst{1,7}
        \and
        \firstname{Teiichiro} \lastname{Matsuzaki}\inst{2}
        \and
        \firstname{Futoshi} \lastname{Minato}\inst{6}
        \and
        \firstname{Rurie} \lastname{Mizuno}\inst{2,8}
        \and
        \firstname{Dai} \lastname{Tomono}\inst{9}
        \and
        \firstname{Yuji} \lastname{Yamaguchi}\inst{3}
}

\institute{
    Department of Advanced Energy Science and Engineering, Kyushu University, Kasuga, Fukuoka 816-8580, Japan.
    \and
    Nishina Center, RIKEN, Wako, Saitama 351-0198, Japan
    \and
    Japan Atomic Energy Agency (JAEA), Tokai, Ibaraki 319-1195, Japan
    \and
    Rutherford Appleton Laboratory, Didcot, Oxfordshire OX11 0QX, United Kingdom
    \and
    High Energy Accelerator Research Organization (KEK), Tokai, Ibaraki 319-1195, Japan
    \and
    Department of Physics, Kyushu University, Fukuoka, Fukuoka 819-0395, Japan
    \and
    Quantum and Spacetime Research Institute, Kyushu University, Fukuoka, Fukuoka 819-0395, Japan
    \and
    TRIUMF, Vancouver, British Columbia, V6T 2A3, Canada
    \and
    Research Center for Nuclear Physics (RCNP), The University of Osaka, Ibaraki, Osaka 567-0047, Japan
}

\abstract{
Negative muon-induced nuclear reactions play a critical role in a wide range of scientific and technological applications; however, comprehensive nuclear data for these processes remain unavailable.
To address this gap, we have launched the Muon Nuclear Data ($\mu$ND) Development Project in Japan, aiming to construct a dedicated data library for muon capture reactions.
The library consists of four sub-libraries: muonic X-ray energies and intensities (XR), lifetimes of muonic atoms and nuclear capture rates (LT), energy spectra of emitted particles (ES), and production branching ratios of residual nuclei (BR).
This project integrates experimental measurements, theoretical modeling, and machine learning techniques to compile and evaluate the data.
We report the current status and recent progress of each sub-library.
}
\maketitle
\section{Introduction}
\label{sec:intro}

A muon is a second-generation charged lepton with a mass approximately 207 times that of an electron.
Muons exist in both positive and negative charge states and decay in vacuum with a mean lifetime of 2.2\,$\mu$s.
The decay modes of positive and negative muons are illustrated below.
In this work, we focus on negative muons which, when implanted in a material, can be captured by atomic nuclei.

Figure 1 schematically illustrates the interaction between a negative muon and an atom.
A negative muon is initially captured into an atomic orbit, forming a muonic atom. Then the captured muon cascades down to the 1s state by emitting Auger electrons and muonic X-rays.
Subsequently, the muon either decays into an electron or is captured by the nucleus via the weak interaction.
The process indicated by the orange arrow, where the muon is captured by the nucleus, in the figure is known as the “muon nuclear capture reaction,” which leads to the formation of a highly excited nucleus.
This nucleus subsequently de-excites through the emission of various particles and gammas.
For muonic atoms of heavy elements, the probability of muon capture exceeds 90\%. 
The excitation energy induced by this reaction typically ranges from 10 to 50 MeV, resulting in the emission of secondary particles and the production of radioactive nuclei.

\begin{figure}[b]
    \centering
    \includegraphics[width=\linewidth]{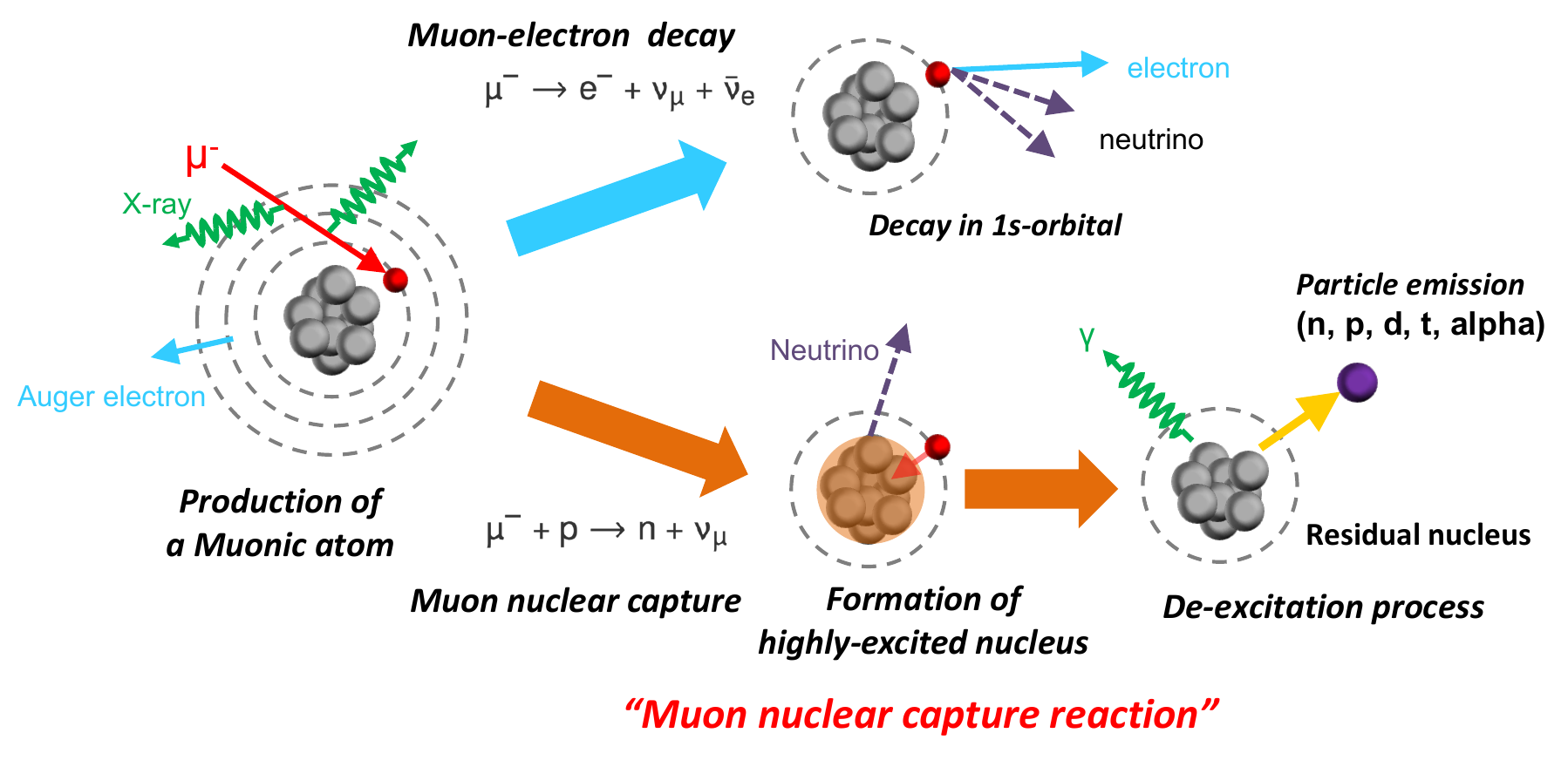}
    \caption{Schematic view of interaction of a negative muon with an atom.}
    \label{fig:placeholder}
\end{figure}

In recent years, muon-nucleus interactions have attracted increasing attention in a wide range of scientific and technological applications, including nuclear physics, nuclear transmutation for waste management, medical radioisotope production, radiation safety in muon facilities, soft-error analysis in modern semiconductor devices, cosmogenic radionuclide production in geoscience, and non-destructive elemental analysis using muon-induced X-ray emission.
Despite these growing demands, a comprehensive and evaluated database for muon capture reactions---hereafter referred to as “muon nuclear data”---has not yet been established.

Under these circumstances, the Muon Nuclear Data ($\mu$ND) Development Project was recently launched in Japan~\cite{Niikura2024-mund}.
The project aims to construct a nuclear data library specifically for muon-induced reaction processes.
In this paper, we present the current status of the project and highlight recent progress in each sub-library.

\section{Outline of Muon Nuclear Data Development Project}
\label{sec:outline}

The muon nuclear data library consists of four sub-libraries:
\begin{itemize}
    \item Muonic X-ray Energies and Intensities (XR)
    \item Lifetimes of Muonic Atoms (LT)
    \item Energy Spectra of Emitted Particles (ES)
    \item Production Branching Ratios of Residual Nuclei (BR)
\end{itemize}
Details of each $\mu$ND sub-library are described in Sect.~\ref{sec:status}.

Similar to conventional neutron nuclear data evaluation, muon nuclear data are evaluated based on experimental measurements and theoretical model calculations.
However, the currently available experimental data alone are insufficient, and additional measurements are required.
In cases where experimental data are scarce, theoretical calculations and machine learning techniques play an essential role in data evaluation.
Through these efforts, reliable datasets are being compiled and stored in experimental databases such as EXFOR, and the evaluated data will ultimately be incorporated into nuclear data libraries such as JENDL.
To promote these activities, the $\mu$ND Working Group was established under the JENDL Committee in 2025.

\section{Current Status and Recent Progress}
\label{sec:status}

This section describes the physical quantities covered in each sub-library and summarizes their current status and recent progress.

\subsection{Muonic X-ray Energies and Intensities (XR)}
\label{sec:xr}

Muonic X-ray spectroscopy provides a powerful method for non-destructive elemental analysis, as the energies of the emitted X-rays are characteristic of each element.
It is also used to determine the absolute nuclear charge radii~\cite{Saito2025-xr}.
To cover the wide energy range of muonic X-rays, Mizuno et al. developed a dedicated detector system~\cite{Mizuno2024}.
In test experiments at the Swiss Muon Source (S$\mu$S) at PSI, both the absolute X-ray intensities and energies for $^{197}$Au and $^{209}$Bi were measured.
These measurements enable in-situ calibration of the detector system over a wide energy range and provide a basis for extending the applicability of muonic X-ray spectroscopy.

Recently, a group at RAL in the UK released the muonic X-ray analysis software ``EVA'' as open-source software~\cite{Bjorklund2026,eva}, improving the data analysis environment.
On the theoretical side, the open-source software ``mudirac'' has been developed to calculate transition energies of muonic atoms~\cite{Sturniolo2021,Liborio2026,mudirac}.
Currently, development of the next-generation code ``mudirac2'' is in progress, with the goal of enabling calculations of not only muonic X-ray energies but also their intensities.

These advances have established an experimental and theoretical foundation for the development of nuclear data for muonic X-ray spectroscopy.

\subsection{Lifetimes of Muonic Atoms / Nuclear Capture Rates (LT)}
\label{sec:lt}

As illustrated in Fig.~1, a muonic atom undergoes two primary processes: ordinary muon decay and muon nuclear capture.
The branching ratio between these two processes can be deduced from the measured lifetime of the muonic atom, which corresponds to the sum of the decay and capture rates.
In general, the capture rate increases with atomic number due to the increased overlap between the muon and nuclear wave functions.

Experimentally, Suzuki et al. performed comprehensive measurements of muonic atom lifetimes at TRIUMF in Canada~\cite{Suzuki1987}. 
Theoretically, Goulard and Primakoff proposed an empirical formula for the capture rate~\cite{Goulard1974}.
These results have been widely used as references in muon capture studies.
Iwamoto et al. proposed a nuclear data evaluation method based on Gaussian process regression, integrating experimental data with theoretical models~\cite{Iwamoto2025}. They also pointed out that the isotope dependence of the muon capture rate remains an open issue due to the lack of experimental data.
In this context, Mizuno et al. conducted lifetime measurements of Si isotopes at the J-PARC MLF, establishing the technical foundation for future systematic measurements~\cite{Mizuno2025-lt}. Building on this progress, further measurements of isotope dependence are planned.

\subsection{Energy Spectra of Emitted Particles (ES)}
\label{sec:es}

The energy spectra of charged particles and neutrons emitted following muon capture provide insight into the time evolution of muon capture reactions, from pre-equilibrium processes to compound nucleus formation and subsequent statistical decays.

For light charged particle measurements, Kawase et al. developed a detector system capable of particle identification over a wide energy range, from a few MeV up to 100 MeV~\cite{Kawase2024}.
Using this system, measurements of charged particles emitted following muon nuclear capture on Si nuclei were carried out at RAL in the UK~\cite{Kawase2026}.
The measured spectra include protons, deuterons, tritons, and alpha particles, and are essential for estimating soft error rates in semiconductor devices.

For neutron measurements, Saito et al. developed a neutron detector array consisting of 21 liquid scintillators, and measured neutron energy spectra for Pd isotopes at RCNP-MuSIC~\cite{Saito2025-es}.
The measured neutron spectra indicate that the energy region around 10 MeV plays an important role in understanding the transition from pre-equilibrium neutron emission to evaporation from the compound nuclei.

\subsection{Production Branching Ratios of Residual Nuclei (BR)}
\label{sec:br}

In recent years, experimental measurements of residual nucleus production following muon capture have been actively progressing. 
Activation methods provide a reliable approach for measuring residual production branching ratios; however, conventional offline activation measurements are generally limited to nuclei with half-lives longer than several minutes.
To overcome this limitation, Niikura et al. developed an in-beam activation method~\cite{Niikura2024} utilizing the time structure of pulsed muon beams at synchrotron facilities.
This technique enables comprehensive measurements of residual nuclei with half-lives as short as several tens of milliseconds. Using this method, residual production branching ratios for various target nuclei have been measured~\cite{Yamaguchi2025, Mizuno2025-br}.

\subsection{Theoretical Model Calculation}
\label{sec:th}

Theoretical model calculations play an essential role in data evaluation when experimental data are scarce.
At the current stage, two complementary approaches are adopted. 
One is based on the models implemented in the general-purpose Monte Carlo transport code PHITS~\cite{Sato2024,Abe2017}, and the other is based on the microscopic and evaporation model (MEM) developed by Minato et al.~\cite{Minato2023}.
Both approaches describe the formation of highly excited compound nuclei following muon capture and their subsequent de-excitation through a combination of theoretical models.

In PHITS, the nuclear excitation energy of the residual nucleus after muon capture is sampled from a phenomenological excitation function proposed by Singer~\cite{Singer1962}.
The pre-equilibrium stage is treated using the JAERI Quantum Molecular Dynamics (JQMD) model~\cite{Niita1995}, followed by statistical decay described by the General Evaporation Model (GEM)~\cite{Furihata2000,Furihata2001}.
To enhance the emission of light composite particles, PHITS incorporates the surface coalescence model (SCM)~\cite{Watanabe2007}.
On the other hand, MEM combines a microscopic nuclear-structure theory based on the second Tamm-Dancoff approximation, a two-component exciton model for the pre-equilibrium emission, and the Hauser–Feshbach model for statistical evaporation process. To better reproduce high-energy proton spectra, meson-exchange currents (MEC)~\cite{Lifshitz1988} are phenomenologically incorporated, enhancing the population of highly excited configurations.

In Ref.~\cite{Kawase2026}, both PHITS and MEM were applied to the analysis of the energy spectra of charged particles produced by muon nuclear capture on Si. Although both models reasonably reproduce proton spectra, noticeable discrepancies are observed for composite particles.
In addition, the experimental branching ratios (BR) and energy spectra obtained in the two experiments~\cite{Kawase2026,Yamaguchi2025,Mizuno2025-br} were compared with the predictions of the two theoretical models.
PHITS successfully reproduces the BRs for charged-particle emission channels; however, it underestimates the high-energy component of the measured spectra. In contrast, MEM reproduces the high-energy region of the charged-particle spectra reasonably well, indicating a significant contribution from MEC. By incorporation of the MEC effect into PHITS, both the branching ratios and the energy spectra are reasonably reproduced.  

\section{Summary and Outlook}
\label{sec:summary}

Accurate muon nuclear data ($\mu$ND) are essential for both fundamental research and technological applications.
To address this need, we have launched a new project to develop a comprehensive $\mu$ND library consisting of four sub-libraries:
(1) Muonic X-ray Energies and Intensities (XR),
(2) Lifetimes of Muonic Atoms / Nuclear Capture Rates (LT),
(3) Energy Spectra of Emitted Particles (ES),
and (4) Production Branching Ratios of Residual Nuclei (BR).
The project integrates experimental, theoretical, and machine learning approaches for the development and evaluation of $\mu$ND library.
In this report, the current status and recent progress in each sub-library were presented.
In the future, we will continue developing and publishing the database based on the research infrastructure established through this project, with the goal of releasing approximately one sub-library per year.

This work was partially supported by JSPS KAKENHI Grant Numbers JP19H05664, JP21H01863, and JP24H00073.

%
% BibTeX or Biber users please use (the style is already called in the class, ensure that the "woc.bst" style is in your local directory)
% \bibliography{paperpile}

\begin{thebibliography}{99}

\bibitem{Niikura2024-mund}
M.~Niikura, S.~Abe, S.~Kawase, T.~Matsuzaki, \mbox{F.~Minato}, R.~Mizuno, Y.~Watanabe, Y.~Yamaguchi, Muon Nuclear Data, in \emph{Proceedings of the Joint Symposium on Nuclear Data and PHITS in 2023}, JAEA-Conf \textbf{2024-002}, pp.~29--34 (2024). \doiwoc{10.11484/jaea-conf-2024-002}

\bibitem{Saito2025-xr}
T.~Y.~Saito, M.~Niikura, T.~Matsuzaki, H.~Sakurai, M.~Igashira, H.~Imao, K.~Ishida, T.~Katabuchi, Y.~Kawashima, M.~K.~Kubo et~al., Muonic x-ray measurement for the nuclear charge distribution: The case of stable palladium isotopes, Phys. Rev. C. \textbf{111}, 034313 (2025). \doiwoc{10.1103/PhysRevC.111.034313}

\bibitem{Mizuno2024}
R.~Mizuno, M.~Niikura, T.~Y.~Saito, T.~Matsuzaki, H.~Sakurai, A.~Amato, S.~Asari, S.~Biswas, I.~Chiu, L.~Gerchow et~al., Development of wide range photon detection system for muonic {X}-ray spectroscopy, Nucl. Instrum. Methods Phys. Res. A \textbf{1060}, 169029 (2024). \doiwoc{10.1016/j.nima.2023.169029}

\bibitem{Bjorklund2026}
M.~Bjorklund, S.~Foxley, T.~Agoro, S.~Basak, J.~Lord, S.~Biswas, K.~Butler, A.~Hillier, EVA: A User-Friendly Analysis Package for Negative
Muon Elemental Analysis, J. Phys.: Conf. Ser. (2026) in press.

\bibitem{eva}
EVA - Elemental Visual Analysis,
\url{https://github.com/ISISMuon/EVA}
(accessed May 2026).

\bibitem{Sturniolo2021}
S.~Sturniolo, A.~Hillier, Mudirac: A dirac equation solver for elemental analysis with muonic x‐rays, X-ray Spectrom. \textbf{50}, 180 (2021). \doiwoc{10.1002/xrs.3212}

\bibitem{Liborio2026}
L.~Liborio, M.~Kumar, S. Devadasan, P.~Jones, M. Plummer, A. Hillier, A. Bartok, MuDirac 1.3.0: A Sustainable Software Tool for Calculating Ground State Nuclear Properties Using Muonic X-Ray Measurements, arXiv:2605.00554 [physics.comp-ph] (2026). \doiwoc{10.48550/arXiv.2605.00554}

\bibitem{mudirac}
MuDirac,
\url{https://github.com/muon-spectroscopy-computational-project/mudirac}
(accessed May 2026).

\bibitem{Suzuki1987}
T.~Suzuki, D.~F.~Measday, J.~P.~Roalsvig, Total nuclear capture rates for negative muons, Phys. Rev. C \textbf{35}, 2212 (1987). \doiwoc{10.1103/PhysRevC.35.2212}

\bibitem{Goulard1974}
B.~Goulard, H.~Primakoff, Nuclear muon-capture sum rules and mean nuclear excitation energies, Phys. Rev. C \textbf{10}, 2034 (1974). \doiwoc{10.1103/PhysRevC.10.2034}

\bibitem{Iwamoto2025}
H.~Iwamoto, M.~Niikura, R.~Mizuno, Comprehensive bayesian machine learning approach to estimating the total nuclear capture rate of a negative muon, Phys. Rev. C. \textbf{111}, 034614 (2025). \doiwoc{10.1103/PhysRevC.111.034614}

\bibitem{Mizuno2025-lt}
R.~Mizuno, M.~Niikura, S.~Akamatsu, T.~Fujiie, K.~Ishida, T.~Ito, T.~Kikuchi, T.~Matsuzaki, F.~Minato, J.~Murata et~al., Lifetime measurement of the muonic atoms of enriched Si isotopes, Phys. Rev. C. \textbf{112}, 024307 (2025). \doiwoc{10.1103/vl7z-rzp8}

\bibitem{Kawase2024}
S.~Kawase, T.~Murota, H.~Fukuda, M.~Oishi, T.~Kawata, K.~Kitafuji, S.~Manabe, Y.~Watanabe, H.~Nishibata, S.~Go et~al., Effect of large-angle incidence on particle identification performance for light-charged ($Z\leq2$) particles by pulse shape analysis with a pad-type {nTD} silicon detector, Nucl. Instrum. Methods Phys. Res. A \textbf{1059}, 168984 (2024). \doiwoc{10.1016/j.nima.2023.168984}

\bibitem{Kawase2026}
S.~Kawase, K.~Kitafuji, T.~Kawata, Y.~Watanabe, M.~Niikura, T.~Matsuzaki, K.~Ishida, R.~Mizuno, D.~Tomono, A.~D.~Hillier et~al., arXiv:2601.09106 [nucl-ex] (2026). \doiwoc{10.48550/arXiv.2601.09106}

\bibitem{Saito2025-es}
T.~Y.~Saito, M.~Niikura, T.~Matsuzaki, S.~Abe, K.~Ishida, S.~Kawase, Y.~Kawashima, T.~Koiwai, K.~Matsui, S.~Momiyama et~al., arXiv:2508.00377 [nucl-ex] (2025). \doiwoc{10.48550/arXiv.2508.00377}

\bibitem{Niikura2024}
M.~Niikura, T.~Y.~Saito, T.~Matsuzaki, K.~Ishida, A.~Hillier, Measurement of the production branching ratios following nuclear muon capture for palladium isotopes using the in-beam activation method, Phys. Rev. C. \textbf{109}, 014328 (2024). \doiwoc{10.1103/PhysRevC.109.014328}

\bibitem{Yamaguchi2025}
Y.~Yamaguchi, M.~Niikura, R.~Mizuno, M.~Tampo, M.~Harada, N.~Kawamura, I.~Umegaki, S.~Takeshita, K.~Haga, Measurement of radionuclide production probabilities in negative muon nuclear capture and validation of Monte Carlo simulation model, Nucl. Instrum. Methods Phys. Res. B \textbf{567}, 165801 (2025). \doiwoc{10.1016/j.nimb.2025.165801}

\bibitem{Mizuno2025-br}
R.~Mizuno, M.~Niikura, T.~Y.~Saito, T.~Matsuzaki, S.~Abe, H.~Fukuda, M.~Hashimoto, A.~D.~Hillier, K.~Ishida, N.~Kawamura et~al., Measurement of production branching ratio after muon nuclear capture reaction of Al and Si isotopes, Phys. Rev. C. \textbf{112}, 054305 (2025). \doiwoc{10.1103/kycz-qprw}

\bibitem{Sato2024}
T.~Sato, Y.~Iwamoto, S.~Hashimoto, T.~Ogawa, \mbox{T.~Furuta}, S.~Abe, T.~Kai, Y.~Matsuya, N.~Matsuda, Y.~Hirata et~al., Recent improvements of the particle and heavy ion transport code system – {PHITS} version 3.33, J. Nucl. Sci. Technol. \textbf{61}, 127 (2024). \doiwoc{10.1080/00223131.2023.2275736}

\bibitem{Abe2017}
S.~Abe, T.~Sato, Implementation of muon interaction models in {PHITS}, J. Nucl. Sci. Technol. \textbf{54}, 101 (2017). \doiwoc{10.1080/00223131.2016.1210043}

\bibitem{Minato2023}
F.~Minato, T.~Naito, O.~Iwamoto, Nuclear many-body effects on particle emission following muon capture on $^{28}$Si and $^{40}$Ca, Phys. Rev. C. \textbf{107}, 054314 (2023). \doiwoc{10.1103/PhysRevC.107.054314}

\bibitem{Singer1962}
P.~Singer, Neutron emission following muon capture in heavy nuclei, Il Nuovo Cimento \textbf{23}, 669 (1962). \doiwoc{10.1007/bf02732735}

\bibitem{Niita1995}
K.~Niita, S.~Chiba, T.~Maruyama, H.~Takada, \mbox{T.~Fukahori}, Y.~Nakahara, A.~Iwamoto, Analysis of the ({N},{xN}$^{'}$) reactions by quantum molecular dynamics plus statistical decay model, Phys. Rev. C \textbf{52}, 2620 (1995). \doiwoc{10.1103/PhysRevC.52.2620}

\bibitem{Furihata2000}
S.~Furihata, Statistical analysis of light fragment production from medium energy proton-induced reactions, Nucl. Instrum. Methods Phys. Res. B \textbf{171}, 251 (2000). \doiwoc{10.1016/s0168-583x(00)00332-3}

\bibitem{Furihata2001}
S.~Furihata, in \emph{Advanced Monte Carlo for Radiation Physics, Particle Transport Simulation and Applications} (Springer Berlin Heidelberg, Berlin, Heidelberg, 2001), pp. 1045--1050.

\bibitem{Watanabe2007}
Y.~Watanabe, D.~N.~Kadrev, Extension of quantum molecular dynamics for production of light complex particles in nucleon-induced reactions, in \emph{ND2007, International Conference on Nuclear Data for Science and Technology}, pp.~1121--1124 (2008). \doiwoc{10.1051/ndata:07196}

\bibitem{Lifshitz1988}
M.~Lifshitz, P.~Singer, Meson-exchange currents and energetic particle emission from $\mu^-$ capture, Nucl. Phys. A \textbf{476}, 684 (1988). \doiwoc{10.1016/0375-9474(88)90330-2}

\end{thebibliography}

% % Format for Journal Reference
% Journal Author, Article title. Journal \textbf{Volume}, page numbers (year). \url{https://doi.org/Article-DOI-number}
% Book Author, \textit{Book title} (Publisher, place, year) page numbers
% \end{thebibliography}

\end{document}